\documentclass[a4,a4wide]{article}
\usepackage[english]{babel}

\usepackage{graphicx}
\usepackage{paralist}

\usepackage{ifthen}
\usepackage{url}
\usepackage{amssymb}
\usepackage{amsmath}
\usepackage{xparse}
\usepackage{amsmath}

\usepackage[usenames,dvipsnames]{xcolor}
\usepackage{xspace}
\usepackage{datatool}
\usepackage{glossaries}

\newcommand{\m}[1]{\ensuremath{#1}\xspace}
\newcommand{\trval}[1]{\m{\mbox{\bf #1}}}

	\newcommand{\limplies}{\m{\Rightarrow}}

	\newcommand{\limpliedby}{\m{\Leftarrow}}
	\newcommand{\limplied}{\limpliedby}
	\newcommand{\lrule}{\m{\leftarrow}}
	\newcommand{\cause}{\m{\stackrel{c}{\lrule}}}
	
	\newcommand{\ltrue}{\trval{t}}
	\newcommand{\lfalse}{\trval{f}}

	\newcommand{\Tr}{\ltrue}

	\newcommand{\voc}{\m{\Sigma}}

	\newcommand{\struct}{\m{I}}
	
	\newcommand{\I}{\m{\mathcal{I}}}

	\newcommand{\theory}{\m{\mathcal{T}}}

	\newcommand{\PP}{\m{\mathcal{P}}}

	\newcommand{\D}{\m{\Delta}}

	\NewDocumentCommand\inter{g+g}{%
	  \IfNoValueTF{#1}
	    {\struct}
	    {\m{#1^{#2}}}}

	\newcommand{\xxx}{\m{\overline{x}}}
	\newcommand{\yyy}{\m{\overline{y}}}
	
	\newcommand{\ddd}{\m{\overline{d}}}

	\newcommand{\ttt}{\m{\overline{t}}}

	\renewcommand{\int}{\m{\mathbb{Z}}}

	\NewDocumentCommand\subs{g+g}{%
	  \IfNoValueTF{#1}
	    {\m{/}}
	    {\m{#1/ #2}}}

	\newcommand{\logicname}[1]{\text{\sc #1}\xspace}

	\newcommand{\foid}{\logicname{FO(\ensuremath{ID})}}
	
	\newcommand{\foidplus}{\logicname{C-Log}} 
	\newcommand{\clog}{\logicname{C-Log}}
	\newcommand{\foclog}{\logicname{FO(\clog)}}

	\newcommand{\fo}{\FO}

\newcommand{\ouracronym}[3]{%
	\newacronym{#1}{#2}{#3}
	\expandafter\newcommand\csname #1\endcsname{\gls{#1}\xspace}%
}
	\ouracronym{FO}{FO}{first-order logic}
	\ouracronym{MX}{MX}{Model Expansion}
	\ouracronym{MO}{MO}{Model Optimization}
	\ouracronym{ASP}{ASP}{Answer Set Programming}
	\ouracronym{TP}{TP}{Theorem Proving}
	\ouracronym{LP}{LP}{Logic Programming}
	\ouracronym{CP}{CP}{Constraint Programming}
	\ouracronym{KR}{KR}{Knowledge Representation}
	\ouracronym{CSP}{CSP}{Constraint Satisfaction Problem}
	\ouracronym{SMT}{SMT}{SAT Modulo Theories}
	\ouracronym{KBS}{KBS}{knowledge-base system}
	\ouracronym{NNF}{NNF}{Negation Normal Form}
	\ouracronym{FNNF}{FNNF}{Flat Negation Normal Form}
	\ouracronym{DefNNF}{DefNNF}{Definition Negation Normal Form}
	\ouracronym{CDCL}{CDCL}{Conflict-Driven Clause-Learning}
	\ouracronym{LCG}{LCG}{Lazy Clause Generation}

	\makeatletter
	\def\ifenv#1{
	\def\@tempa{#1}%
	\def\@ttempa{#1*}%
	\ifx\@tempa\@currenvir
	\expandafter\@firstoftwo
	\else
	\expandafter\@secondoftwo
	\fi
	}
	\makeatother

	\newcommand{\ddrule}[4]{\ensuremath{#1 \leftarrow #2 & \{#3\} & #4}}
	\newcommand{\drule}[2]{\ensuremath{#1 & \leftarrow & #2}}

	\newcommand{\darule}[4]{\ensuremath{#1 \leftarrow #2 & \{#3\} & #4}}
	\newcommand{\arule}[2]{\ensuremath{#1 \, &\leftarrow \, #2}}

	\newenvironment{ldef}{\left\{\begin{array}{l@{ \,}l@{\,}l}}{\end{array}\right\}}
	\newenvironment{ltheo}{\[\begin{array}{l}}{\end{array}\]\ignorespacesafterend}

	\newcommand{\LNDRule}[2]{
	\ifenv{array}
	{\drule{#1}{#2}}
	{ \ifenv{align}
		{\arule{#1}{#2}}
		{\ifenv{align*}
		{\arule{#1}{#2}}
		{ERROR: using LDRule in unsupported environment: \@currenvir}
		}
	}
	}

	\newcommand{\LDRule}[4]{
	\ifenv{array}
	{\ddrule{#1}{#2}{#3}{#4}}
	{ \ifenv{align}
		{\darule{#1}{#2}{#3}{#4}}
		{\ifenv{align*}
		{\darule{#1}{#2}{#3}{#4}}
		{ERROR: using LDRule in unsupported environment: \@currenvir}
		}
	}
	}

	\NewDocumentCommand\LRule{m+g+g+g}{%
		\IfNoValueTF{#2}%
		{#1.&}{%
		\IfNoValueTF{#3}
		{\LNDRule{#1}{#2.}}
		{\LDRule{#1}{#2.}{#3}{#4}}%
		}
	}

	\NewDocumentCommand\CLRule{m+g}{%
	\ifenv{array}
	{\cdrule{#1}{#2}}
	{ \ifenv{align}
		{\carule{#1}{#2}}
		{\ifenv{align*}
			{\carule{#1}{#2}}
			{ERROR: using CLRule in unsupported environment: \@currenvir}
		}
	}
	}

	\NewDocumentCommand\carule{m+g}{%
		\IfNoValueTF{#2}
			{\ensuremath{#1.}}
			{\ensuremath{#1 \, &\cause \, #2}}}
	\NewDocumentCommand\cdrule{m+g}{%
		\IfNoValueTF{#2}
			{\ensuremath{#1.}}
			{\ensuremath{#1 & \cause & #2}}}

	\newcommand{\algrule}[4]{
	\hbox{{#1}:}& 
	\quad #2 ~\longrightarrow~ #3 
	\hbox{~ if } #4\\
	}

	\newcommand{\AlgoRule}[4]{
	\ifenv{array}
	{\algrule{#1}{#2}{#3}{#4}}
		{ERROR: using AlgoRule in unsupported environment: \@currenvir}
	}

\newcommand{\commentstyle}{\color{Gray}}

	\usepackage[final]{listings}

	\lstdefinelanguage{idp}{
		morekeywords=[1]{namespace,vocabulary,theory,structure,procedure,term,set,formula, spec, specification},
		morekeywords=[2]{include,using,type,isa,contains,partial,extern,LFD,GFD,constructed,from,constraint,func,pred,supertype,of,subtype,define},
		morekeywords=[3]{int,float,char,string,nat},
		morekeywords=[4]{if,then,else,for,end},
		morecomment=[s]{/*}{*/},	
		morecomment=[l]{//}
	}
	\newcommand{\ignore}[1]{}

	\newboolean{nocomments}
	\setboolean{nocomments}{false}

	\newboolean{commentmargin}
	\setboolean{commentmargin}{true}

	\newcommand{\namedcomment}[3]{
		\ifthenelse{\boolean{nocomments}}
		{} 
		{ 
			\ifthenelse{\boolean{commentmargin}}
				{ {\color{#3} \marginpar{\color{#3}\sc #2}#1}  } 
				{  {\color{#3} {\sc #2}: #1}  } 
		}
	}
	\newcommand{\mnamedcomment}[3]{\ifthenelse{\boolean{nocomments}}{}{{\marginpar{ \color{#3}{\sc #2}:#1}}}}

	\newcommand{\todo}[1]{\namedcomment{#1}{TODO}{blue}}

	\newcommand{\bart}[1]{\namedcomment{#1}{bb}{red}}
	
	\newcommand{\marc}[1]{\namedcomment{#1}{md}{orange}}

	\usepackage{soul}

\newcommand{\keyword}[2]{%
	\expandafter\newcommand\csname #1\endcsname{#2\xspace}%
	\expandafter\newcommand\csname #1s\endcsname{#2s\xspace}%
	\expandafter\newcommand\csname #1ness\endcsname{#2ness\xspace}%
}

\usepackage{etoolbox}

\newcommand\setcitation[2]{%
  \csdef{mycommoncitation#1}{#2}}
\newcommand\getcitation[1]{%
  \csuse{mycommoncitation#1}}

\setcitation{IDP}{WarrenBook/DeCatBBD14}
\setcitation{fodot}{tocl/DeneckerT08}
\setcitation{CP}{Apt03}
\setcitation{EZCSP}{lpnmr/Balduccini11}
\setcitation{KR}{Baral:2003}
\setcitation{ASPComp2}{lpnmr/DeneckerVBGT09}
\setcitation{ASPComp3}{LPNRM/Calimeri11}
\setcitation{ASPComp4}{conf/lpnmr/AlvianoCCDDIKKOPPRRSSSWX13}
\setcitation{CPSupport}{ictai/DeCat13}
\setcitation{functionDetection}{iclp/DeCatB13}
\setcitation{fodot2asp}{Denecker12}
\setcitation{Inca}{aaai/DrescherW11}
\setcitation{csp2asp}{ijcai/DrescherW11}
\setcitation{DPLLT}{cav/GanzingerHNOT04}
\setcitation{AspInPractice}{synthesis/2012Gebser}
\setcitation{clasp}{ai/GebserKS12}
\setcitation{oclingo}{kr/GebserGKOSS12}
\setcitation{gringo}{lpnmr/GebserST07}
\setcitation{cmodels}{aaai/GiunchigliaLM04}
\setcitation{inputster}{tplp/Jansen13}
\setcitation{DLV}{tocl/LeonePFEGPS06}
\setcitation{LearningPaper}{corr/Blockeel13}
\setcitation{clog}{iclp/BogaertsVDV14}
\setcitation{inferenceClog}{ecai/BogaertsVDV14}
\setcitation{AFT}{DeneckerBV12}
\setcitation{}{}
\setcitation{}{}
\setcitation{}{}
\setcitation{}{}
\setcitation{}{}
\setcitation{}{}
\setcitation{}{}
\setcitation{}{}
\setcitation{}{}
\setcitation{}{}
  
\newcommand\mycite[1]{%
      \ifcsname mycommoncitation#1\endcsname%
   \cite{\getcitation{#1}}%
  \else%
    {\color{blue}ERROR: undefined citation key ``#1''}
  \fi%
}

\usepackage{amsthm}

\newcommand{\univBQ}[2]{\m{\forall #1 [#2]}}
\newcommand{\exisBQ}[2]{\m{\exists #1 [#2]}}

\newcommand{\pred}[1]{\m{\mathrm{#1}}}
\renewcommand\pred[1]{\m{#1}}

\NewDocumentCommand\lattice{g+g}{%
  \IfNoValueTF{#1}
    {\m{L}}
    {\m{\left(#1,#2\right)}}%
}

\newcommand{\Bilat}[1]{\m{#1^2}}
\NewDocumentCommand\bilat{g}{%
  \IfNoValueTF{#1}
    {\Bilat{\lattice}}
    {\Bilat{#1}}%
}

\newcommand{\cBilat}[1]{\m{#1^c}}
\NewDocumentCommand\cbilat{g}{%
  \IfNoValueTF{#1}
    {\cBilat{\lattice}}
    {\cBilat{#1}}%
}

\newcommand{\Operator}{\m{O}}
\NewDocumentCommand\Op{g}{%
  \IfNoValueTF{#1}
    {\m{\Operator}}
    {\m{\Operator \left(#1\right)}}%
}

\newcommand{\Approximator}{\m{A}}
\NewDocumentCommand\Ap{g}{%
  \IfNoValueTF{#1}
    {\m{\Approximator}}
    {\m{\Approximator \left(#1\right)}}%
}

\newcommand{\calo}{\m{{I_o}}}

\NewDocumentCommand\struclat{g}{%
	\IfNoValueTF{#1}
		{\m{\lattice^\voc_\calo}}
		{\m{\lattice^\voc_{#1}}}
}
\NewDocumentCommand\domstruclat{g}{%
	\IfNoValueTF{#1}
		{\m{\lattice^\voc_\calo}}
		{\m{\lattice^\voc_{#1}}}
}

\NewDocumentCommand\proj{g}{%
	\IfNoValueTF{#1}
		{\m{p}}
		{\m{p(#1)}}
}

\newcommand{\valueofs}[1]{\m{\mathrm{eff}_{\struct,\ch}}}

\newacronym{CNF}{CNF}{Causal Normal Form}

\keyword{naming}{naming}
\keyword{Naming}{Naming}
\keyword{Chfun}{$\D$-selection}
\keyword{chfun}{$\D$-selection}
\keyword{Chpfun}{$\relat{\D}$-selection}
\keyword{chpfun}{$\relat{\D}$-selection}
\keyword{welldef}{well-defined}
\newcommand{\ch}{\m{\zeta}}

\usepackage{amsthm}

\theoremstyle{plain}
\newtheorem{thm}{Theorem}[section]

\newtheorem{definition}[thm]{Definition}

\theoremstyle{definition}

\newtheorem{example}[thm]{Example}

\newtheorem{ex*}{Example}
\newcommand{\cont}{continued}

\newtheoremstyle{example_contd}
{0.5\topsep} 
{0.5\topsep} 
{\normalfont}
{0pt}
{\bfseries}
{.}
{5pt plus 1pt minus 1pt}
{\thmname{#1} \thmnumber{ #2}\thmnote{#3} (\cont)}

\newtheoremstyle{plain_contd}
{0.5\topsep} 
{0.5\topsep} 
{\itshape}
{0pt}
{\bfseries}
{.}
{5pt plus 1pt minus 1pt}
{\thmname{#1} \thmnumber{ #2}\thmnote{#3} (\cont)}

\theoremstyle{example_contd} 

\theoremstyle{plain_contd} 

\theoremstyle{plain}

\newcommand{\aspnot}{\m{\text{not}\,}}
\newcommand{\asprule}{\m{\,\text{:-}\,}}
\newcommand{\M}{\m{\mathcal{M}}}

\RenewDocumentCommand\CLRule{m+g}{%
		\IfNoValueTF{#2}
			{\m{& #1.}}
			{\m{& #1 \cause #2 }}}

\lstdefinelanguage{br}{
		morekeywords=[1]{if,then,else,for,end,while,when,rule},
	}
	\newcommand{\Catom}[1]{#1}
	\NewDocumentCommand\Call{g+g+g}{%
	\IfNoValueTF{#1}%
	    {\m{\mathbf{All}}}%
	    {\m{\mathbf{All\,}#1[#2]: #3}}%
	 }

	\NewDocumentCommand\Cand{g+g}{%
	\IfNoValueTF{#1}%
	    {\m{\mathbf{And}}}%
	    {\m{#1\mathbf{\,And\,}#2}}%
	}
	\NewDocumentCommand\Csel{g+g+g+g}{%
	\IfNoValueTF{#1}%
	    {\m{\mathbf{Select}}}%
		{\IfNoValueTF{#4}%
			{\m{\mathbf{Select\,}#1[#2]: #3}}%
			{\m{\mathbf{Select}_{#1}\,#2[#3]: #4}}%
	}}
	\NewDocumentCommand\Cor{g+g+g+g+g}{%
	\IfNoValueTF{#1}%
		{\m{\mathbf{Or}}}%
		{\m{#1\mathbf{\,Or\,}#2}%
			\IfValueTF{#3}{\mathbf{\,Or\,}#3%
				\IfValueTF{#4}{\mathbf{\,Or\,}#4%
					\IfValueTF{#5}{\mathbf{\,Or\,}#5}{}%
				}{}%
			}{}%
		}%
	}
	\NewDocumentCommand\Cnew{g+g+g}{%
	  \IfNoValueTF{#1}%
	    {\m{\mathbf{New}}}%
	    {\IfNoValueTF{#3}%
			{\m{\mathbf{New\,}#1: #2}}%
			{\m{\mathbf{New\,}#1[#2]: #3}}}}%

	\NewDocumentCommand\Cif{g+g}{%
	\IfNoValueTF{#1}%
	    {rule}%
	    {\m{#2 \leftarrow #1 }}%
	}

\keyword{CEE}{CEE}

\renewcommand{\CLRule}[2]{\Cif{#2}{#1}}

\renewcommand{\subs}[2]{#2:#1}

\renewcommand{\struct}{\m{J}}

\newcommand{\tcaused}{endogenous\xspace}
\newcommand{\topen}{exogenous\xspace}

\newcommand{\xemph}[1]{\emph{#1}} 
\renewcommand{\xemph}[1]{\emph{#1}\xspace }

\keyword{cset}{effect set} 
\keyword{poscset}{possible effect set} 

\setboolean{commentmargin}{false}

\renewcommand{\P}{\pred{P}}
\newcommand{\TB}{\pred{T}(\pred{B})}
\newcommand{\TS}{\pred{T}(\pred{S})}

\Catom

\newcommand{\relat}[1]{\m{rel(#1)}}

\keyword{nestingdepth}{nesting depth}
\keyword{Nestingdepth}{Nesting depth}

 \ouracronym{nestnf}{NestNF}{Nesting Normal Form} 
\ouracronym{deff}{DefF}{Definition Form}

\renewcommand\foclog{\logicname{FO(C)}}

\usepackage{aaai}
\usepackage{times}
\usepackage{helvet}
\usepackage{courier}
\title{\foclog and Related Modelling Paradigms}
\author{Bart Bogaerts \and Joost Vennekens \and Marc Denecker\\
Department of Computer Science, KU Leuven \\
\{bart.bogaerts, joost.vennekens, marc.denecker\}@cs.kuleuven.be\\
\AND
Jan Van den Bussche\\
Hasselt University \& transnational University of Limburg\\
 jan.vandenbussche@uhasselt.be}

\begin{document}
	\nocopyright

\maketitle
 \bibliographystyle{aaai}

\setboolean{nocomments}{true}

\begin{abstract}
Recently, \clog was introduced as a language for modelling causal processes.
Its formal semantics has been defined,
but the study of this language is far from finished. 
In this paper, we compare \clog to other declarative modelling languages. 
More specifically, we compare to \fo, and argue that \clog and \fo are orthogonal and that their integration, \foclog, is a knowledge representation language that allows for clear and succinct models. 
We compare \foclog to E-disjunctive logic programming with the stable semantics, and define a fragment on which both semantics coincide.
Furthermore, we discuss object-creation in \foclog, relating it to mathematics, business rules systems, and data base systems.

\end{abstract}

\section{Introduction}\label{sec:intro}
Previous work introduced \clog \mycite{clog}, an expressive language construct to describe causal processes, and \foclog, its integration with classical logic. 
In that work, it is indicated that \clog shows similarities to many other languages and it is
suggested that \clog could serve as a tool to study the semantical relationship between these languages. 
In this paper, we take the first steps for such a study: we discuss the relationship of \foclog with other paradigms and through this discussion, provide a comprehensive overview of the informal semantics of \foclog. 

\clog and \fo are syntactically very similar, but semantically very different languages. 
In this paper we formalise the semantical relationship between \clog and \fo, and argue how their integration, \foclog, is a rich language in which knowledge can be represented succinctly and clearly. 

We explain how modelling in \foclog relates to the ``generate, define, and test'' methodology used in answer set programming.
We discuss how \foclog relates to disjunctive logic programs with existential quantification in rule heads \cite{tplp/YouZZ13}, both informally and formally,
and we identify a subset of E-disjunctive logic programs on which stable semantics corresponds to the \foclog semantics.
We also discuss four important knowledge representation constructs that \foclog adds with respect to E-disjunctive logic programs: \emph{nested rules} (in fact, arbitrary nesting of expressions), \emph{dynamic choice}, \emph{object creation}, and \emph{a more modular semantics}.

Furthermore, we discuss object-creation in related paradigms. 
One of those discussed paradigms is the field of deductive databases, where extensions of Datalog have been defined. 
In~\cite{jcss/AbiteboulV91}, rules with existentially quantified head variables are used for object creation. 
It is remarkable to see how the same extension of logic programs is used sometimes (e.g., in~\cite{tplp/YouZZ13}) for selection, and sometimes (e.g., in~\cite{jcss/AbiteboulV91}) for object-creation. 
Consider for example a rule 
\[\forall X: \exists Y: P(X,Y) \asprule q(X).\]
Viewing this rule as a rule in an E-disjunctive logic program, it corresponds to the \clog expression
\[\Call{ X}{q(X)}{\Csel{Y}{\ltrue}{P(X,Y)}},\]
where for every $X$ satisfying $q$, one existing value $Y$ is selected, and $P(X,Y)$ is caused. The selected $Y$ can be different or equal for different $X$'s. 
On the other hand, in case this same rule occurs in a LogicBlox~\cite{datalog/GreenAK12} specification,  it corresponds to the \clog expression
\[\Call{ X}{q(X)}{\Cnew{Y}{P(X,Y)}},\]
where for every $X$ satisfying $q$ a new value $Y$ is invented. Thus implying among others that all of these values are different. 
The explicit  distinction \clog makes between object-creation and selection is necessary for studying the relationship between these languages.

The rest of this paper is structured as follows. 
In Section \ref{sec:prelims} we give  preliminaries, including the syntax and informal semantics of \clog. In Sections \ref{sec:fo} and \ref{sec:asp}, we focus on the creation-free fragment of \clog, i.e., on expressions without the \Cnew-operator:
first,  we compare \clog to \fo and discuss the integration of these two; afterwards,	 we compare \clog to E-disjunctive logic programs. 
In Section \ref{sec:new}, we discuss object-creation in \clog by providing simple intuitive examples  and relating the \Cnew-operator to other languages with similar forms of object-creation. We conclude in Section \ref{sec:concl}.

\marc{intro zoals ze nu is is helemaal niet goed: wat doen die formules met select en new hier al? Wat moet er staan? New modelling formals relation to CWA, complex formal semantics, but natural informal semantics. Zie ook brainstor mail die nu aan het gestuurd worden is...}
\section{\clog}\label{sec:prelims}
We assume familiarity with the basics of first-order logic. 
Vocabularies, formulas, and terms are defined as usual.
We use \ltrue for truth and \lfalse for falsity. 
$\sigma^\I$ denotes
the interpretation of symbol $\sigma $ in structure \I.
\emph{Domain atoms} are atoms of the form $P(\ddd)$ where the $d_i$ are domain elements.
We use restricted quantifications \cite{PreyerP02}, e.g., in \FO, these are formulas of the form
$\univBQ{x}{\psi}: \varphi\label{formula:BinQuantForall}$ or $\exisBQ{x}{\psi}: \varphi\label{formula:BinQuantExists}$,
meaning that $\varphi$ holds for all (resp.~for a) $x$ such that $\psi$ holds.
The above expressions are syntactic sugar for
$	\forall x: \psi \limplies \varphi
$ and $ 
	\exists x: \psi \land \varphi,
$
but such a reduction is not possible for other restricted quantifiers in \clog.
We call $\psi$ the \emph{qualification} and $\varphi$ the \emph{assertion} of the restricted quantifications.
From now on, let \voc be a relational vocabulary, i.e., \voc consists only of predicate, constant and variable symbols.

In what follows we briefly repeat the syntax and informal semantics of \clog. For more details and an extensive overview of the formal semantics of \clog,  we refer to \mycite{clog}.

\subsection{Syntax of \clog}
\begin{definition}
 \emph{Causal effect expressions} (\CEE) are defined inductively as follows:
\begin{itemize}
	\item if $P(\ttt)$ is an atom, then $\Catom{P(\ttt)}$ is a \CEE,
	\item if $\varphi$ is an \fo formula and $C'$ is a \CEE, then $\Cif{\varphi}{C'}$ is a \CEE,
	\item if $C_1$ and $C_2$ are \CEEs, then $\Cand{C_1}{C_2}$ is a \CEE,
	\item if $C_1$ and $C_2$ are \CEEs, then $\Cor{C_1}{C_2}$  is a \CEE,
	\item if $x$ is a variable, $\varphi$ is a first-order formula and $C'$ is a \CEE, then $\Call{x}{\varphi}{C'}$  is a \CEE,
	\item if $x$ is a variable, $\varphi$ is a first-order formula and $C'$ is a \CEE, then $\Csel{x}{\varphi}{C'}$ is a \CEE,
	\item if $x$ is a variable and $C'$ is a \CEE, then $\Cnew{x}{C'}$ is a \CEE.
\end{itemize}
\end{definition}

We call a \CEE an \emph{atom-expression}  (respectively \emph{\Cif-},   \emph{\Cand-}, \mbox{\emph{\Cor-},}
\emph{\Call-}, \emph{\Csel-} or \emph{\Cnew-expression}) if it is of the corresponding form. 
We use $\Call{\xxx}{\varphi}{C}$ as an abbreviation for $\Call{x_1}{\ltrue}{\dots \Call{x_n}{\varphi}{C}}$ and similar for \Csel-expressions.
We call a predicate symbol $P$ \xemph{\tcaused}in $C$ if $P$ occurs as the symbol of a (possibly nested) atom-expression in $C$, i.e., 
if $P$ occurs in $C$ but not only in first-order formulas. 
All other symbols are called \xemph{\topen}in $C$. 
An occurrence of a variable $x$ is \emph{bound} in a \CEE if it occurs in the scope of a quantification over that variable 
($\forall x$, $\exists x$, $\Call\, x$, $\Csel\, x$, or $\Cnew\, x$) and \emph{free} otherwise. A variable is \emph{free} in a \CEE if it has free occurrences. 
A \emph{causal theory}, or \emph{\foidplus theory} is a \CEE without free variables. 
We often represent a causal theory as a set of \CEEs; the intended causal theory is the \Cand-conjunction of these \CEEs.

\subsection{Informal Semantics of \clog}
In this section, we discuss the informal semantics of \CEEs. 
We repeat the driving principles on a simple example---one without non-determinism---and discuss more complex expressions afterwards. 

\subsubsection{Driving Principles}
Following the philosophy of \cite{journal/tplp/VennekensDB10}, the semantics of \clog is based on two principles that are common in causal modelling. 
The first is the distinction between {\em endogenous} and {\em exogenous} properties, 
i.e., those whose value is determined by the causal laws in the model and those whose value is not, respectively \cite{Pearl00}. 
The second is the {\em default-deviant} assumption, used also by, e.g., \cite{Hall04,Hitchcock07}. 
The idea here is to 
assume that each endogenous property of the domain has some ``natural'' state, that it will be in whenever nothing is acting upon it. 
For ease of notation, \clog identifies the default state with falsity, and the deviant state with truth.
For example, consider the following simplified model of a bicycle, in which a pair of gear wheels can be put in motion by pedalling:
\begin{align}
\pred{Turn}(\pred{BigGear})   &\leftarrow \pred{Pedal}.         \label{pedal} \\
\pred{Turn}(\pred{BigGear})   &\leftarrow \pred{Turn}(\pred{SmallGear}).\label{bigg}   \\
\pred{Turn}(\pred{SmallGear}) &\leftarrow \pred{Turn}(\pred{BigGear}).  \label{smallg}
\end{align}
Here, $\pred{Pedal}$ is exogenous, while $\pred{Turn}(\pred{BigGear})$ and $\pred{Turn}(\pred{SmallGear})$ are endogenous. The semantics of this causal model is given by a straightforward ``execution'' of the rules. The domain starts out in an initial state, in which all endogenous atoms have their default value {\em false} and the exogenous atom $\pred{Pedal}$ has some fixed value. If $\pred{Pedal}$ is true, then the first rule is applicable and may be fired (``$\pred{Pedal}$ causes $\pred{Turn}(\pred{BigGear})$'') to produce a new state of the domain in which $\pred{Turn}(\pred{BigGear})$ now has its deviant value {\em true}. In this way, we construct the following sequence of states (we abbreviate symbols by their first letter):
\begin{equation}\label{branch}
 \{\P\} \rightarrow \{\P,\TB\} \rightarrow \{\P,\TB,\TS\}
\end{equation}
In general, given a causal theory \D, a causal process is a (possibly transfinite) sequence of intermediate states, starting from the default state   such that, at each state, the effects described by $\D$ take place.
This notion of causal process is based on the following principles:
\begin{itemize}
\item The principle of \emph{sufficient causation} states that if the precondition to a causal
law is satisfied, then the event that it triggers must eventually happen. For example, the process described in (\ref{branch}) cannot stop after the first step: there is a cause for $\pred{Turn}(\pred{SmallGear})$, hence this should eventually happen.
 \item The principle of \emph{universal causation} states that all changes to the state of the
domain must be triggered by a causal law whose precondition is satisfied. For example, the small gear can only turn if the big gear turns.
\item The principle of \emph{no self-causation} states that nothing can happen based on itself. E.g., if rule (\ref{pedal}) would be excluded from the causal theory, the gears cannot start rotating by themselves.  
\end{itemize}

\subsubsection{Complex Expressions}

A (possibly infinite) structure is a model of a causal theory \D if it is the final state of a (non-deterministic) causal processes described by \D.  
In order to define these processes correctly, one should know 
the events that take place in every state. 
We call the set of those events the \xemph{\cset}of the causal theory.
There are two kinds of effects that can be described by a  causal theory: 1) flipping an atom from its default to its deviant state
and 2) creating a new domain element.
We now explain in a compositional way what the \cset  of a  causal theory is in a given state of affairs, which we represent as usual by a structure.

The effect of an atom-expression $A$ is that $A$ is flipped to its deviant state.
A conditional effect, i.e., a rule expression, causes the \cset  of its head if its body is satisfied in the current state, and   nothing otherwise.
The \cset described by an \Cand-expression is the union of the \csets of its two subexpressions;  an \Call-expression $\Call{x}{\varphi}{C'}$ causes 
the union of all \csets of $C'(x)$ for those $x$'s that satisfy $\varphi$.
An expression $\Cor{C_1}{C_2}$ non-deterministically causes either the \cset of $C_1$ or the \cset of $C_2$; a \Csel-expression $\Csel{x}{\varphi}{C'}$ causes the \cset 
of $C'$ for a non-deterministically chosen   $x$ that satisfies $\varphi$. 
An object-creating \CEE $\Cnew{x}{C'}$ causes the creation of a new domain element $n$ and  the \cset of $C'(n)$.

Informally, \CEEs only cause changes to the state once (for each of its instantiations), e.g., a \Csel-expression $\Csel{x}{\varphi}{C'}$ causes the \cset 
of $C'$ for a non-deterministically chosen   $x$ once, and cannot cause $C'$ for another $x$ afterwards.

\begin{example}\label{ex:american}\label{ex:lottery}
	Permanent residence in the United States can be obtained in several ways. 
	One way is  passing the naturalisation test.
	Another way is by playing the ``Green Card Lottery'', where each year a number of lucky winners are  randomly selected and granted permanent residence.
	We model this as follows:
	\begin{align*} \begin{ldef}
		&\Call{p}{\pred{Apply}(p) \land \pred{PassedTest}(p)}{ \pred{PermRes}(p)}\\
		&\CLRule{(\Csel{p}{\pred{Play}(p)}{\pred{PermRes}(p)})}{\pred{Lottery}.}\end{ldef}
	\end{align*}
	The first \CEE describes the ``normal'' way to obtain permanent residence; 
	the second rule expresses that one winner is selected among everyone who plays the lottery. 
If \I is a structure in which $\pred{Lottery}$ holds, due to the non-determinism, there are many \poscsets of the above \CEE, namely the sets $\{\pred{PermRes}(p) \mid p\in \inter{\pred{Apply}}{\I} \land p\in\inter{\pred{PassedTest}}{\I} \}$ $ \cup$ $\{\pred{PermRes}(d)\}$ for some  $d\in\inter{\pred{Play}}{\I}$. 

Models of this causal theory are structures such that everyone who applies and passes the test has permanent residence, and in case the lottery happens, one random person who played the lottery as well, and such that furthermore no-one else obtains permanent residence. The principle of sufficient causation guarantees a form of closed world assumption: you can only obtain residence if there is a rule that causes you to obtain this nationality.
The two  \CEEs are considered independent: the winner could be one of the people that obtained it through standard application, as well as someone else, i.e., the semantics allows both minimal and non-minimal models.

Note that in the above, there is a great asymmetry between $\pred{Play}(p)$, which occurs as a qualification of \Csel-expression, 
and  $\pred{PermRes}(p)$, which occurs as a caused atom.  This means that the effect will never cause atoms of the form $\pred{Play}(p)$, but only atoms of the form $\pred{PermRes}(p)$. 
This is one of the cases where the qualification of an expression cannot simply be eliminated. 
\end{example}

\begin{example}\label{ex:mail}
Hitting the ``send'' button in your mail application causes the creation of a new package containing a specific mail. 
That package is put on a channel and will be received some (unknown) time later. 
As long as the package is not received, it stays on the channel.
In \foidplus, we model this as follows:
\begin{align*}\begin{ldef}
	&\Call{m,t}{\pred{Mail}(m)\land \pred{HitSend}(m,t)}{\Cnew{p}{  \\
	  &\quad    \Cand{\pred{Pack}(p)}{\Cand{\pred{Cont}(p,m)}{\Cand{\pred{OnCh}(p,t+1)}{\\
	  &\quad \Csel{d}{d>0}{\pred{Received}(p,t+d)}                 }}}              }          }\\
	&\Call{p, t}{\pred{Pack}(p) \land \pred{OnCh}(p,t) \land \lnot \pred{Received}(p,t)}{\\
	&\quad \pred{OnCh}(p,t+1)}
\end{ldef}
\end{align*}
Suppose an interpretation $\inter{\pred{HitSend}}{\I}=\{(\pred{MyMail},0)\}$ is given.  
A causal process then unfolds as follows: it starts in the initial state, where all \tcaused predicates are false.
The \cset of the above causal effect in that state consists of 1) the creation of one new domain element, say $\_p$, and 2) the caused atoms $\pred{Pack}(\_p)$, $\pred{Cont}(\_p,\pred{MyMail})$, $\pred{OnCh}(\_p,1)$ and $\pred{Received}(\_p,7)$, 
where instead of $7$, we could have chosen any number greater than zero. 
Next, it continues, and in every step $t$, before receiving the package, an extra atom $\pred{OnCh}(p,t+1)$ is caused. 
Finally, in the seventh step, no more atoms are caused; the causal process ends. 
The final state is a model of the causal theory.
\end{example}

\subsection{\foclog}
First-order logic  and  \foidplus have a straightforward integration, \foclog. Theories in this logic are sets of FO sentences and causal theories. A model of such a theory is a structure that is a model of each of its expressions (of each of its \CEEs and sentences).  An illustration is  the mail protocol from  Example~\ref{ex:mail}, which we can extend with the ``observation''  that at some time, two packages are on the channel:
\[ \exists t, p_1, p_2 [p_1\neq p_2]: \pred{OnCh}(p_1,t) \land \pred{OnCh}(p_2,t).\] 
 Models of this theory represent states of
affairs where at least once two packages are on the channel simultaneously.  This entirely differs from \Cand-conjoining our \CEE with
\begin{align*}
 & \Csel{t,p_1,p_2}{ p_1 \neq p_2}{\Cand{\pred{OnCh}(p_1,t)}{ \pred{OnCh}(p_2,t)}}.
\end{align*}
The resulting \CEE would have unintended models in which  two  packages suddenly appear on the channel for no reason. 
Note that in the definitions of \clog, we restricted attention to relational vocabularies. All the theory can straightforwardly be generalised as long as function symbols do not occur as endogenous symbols in \CEEs, i.e., if they only occur in \fo sentences or as exogenous symbols in causal theories. 

\section{\clog, FO, and \foclog}\label{sec:fo}
\marc{hier moet een stuk komen over causale versus assertionele info}
There is an obvious syntactical correspondence between \fo and creation-free \clog (\clog without \Cnew-expressions):
$\Cand$ corresponds to $\land$,  $\Cor$ to $\lor$,  $\lrule$ to $\limplied$, $\Call$ to $\forall$, and $\Csel$ to $\exists$.
As already mentioned above, expressions in \clog have an entirely different meaning than the corresponding \fo expression. 
A \clog expression describes a process in which more and more facts are  caused, while an \fo expression describes a truth. 
For example
$ \Cor{P}{Q}$ \marc{dit is geen goed vb: P en Q betekenen niets en dit kan de indruk geven een exclusieve of te zijn...}
describes a process that picks either $P$ or $Q$ and makes one of them true, hence its models are structures in which exactly one of the two holds.
On the other hand, the \fo sentence 
$P\lor Q$
has more models, namely also one  in which both hold. 
We generalise this observation:
\begin{thm}
  Let $\D$ be a creation-free causal theory over \voc and $\theory_\D$ the corresponding \fo theory (the theory obtained from \D by replacing \Call by $\forall$, \Csel by $\exists$, $\Cor$ by $\lor$, $\Cand$ by $\land$, and $\lrule$ by $\limplied$).
 Then for every \voc-structure \I, if $\I\models\D$, then also $\I\models\theory_\D$. 
\end{thm}

The reverse often does not hold: there is no obvious way to translate any \fo formula to a \clog expression.
In some cases, it is possible to find an inverse transformation, for example for positive (negation-free) \fo theories.
This would yield a constructive way to create models for a positive \fo theory, which is not a surprising, nor a very interesting result; another constructive way to get a model of such a theory would be to make everything true.
But it is interesting to view \clog theories as a constructive way to create a certain structure. \marc{dat hele stuk over constructive moet weg}
This shows that modelling in \clog is orthogonal to modelling in \fo. 
In \fo, by default everything is open, every atom can be true or false arbitrarily. 
Every constraint removes worlds from the set of possible worlds. 
In \clog on the other hand, all \tcaused symbols are by default false. 
Adding extra rules to a \clog theory can result in more models (when introducing extra non-determinism), or modify worlds.
In some cases, one of the approaches is more natural than the other. 

\marc{goed vb maar wat doet dit hier?}\marc{voorbeeld moet herschreven worden: eerst in NL het probleem schetsen, dan bespreken: in FO is dit moeilijk en tedious werk om te krijgen. Dan: in \clog gemakkelijk. Wat is het verhaal? Zeer eenvoudige theorie: geen recursie, heel natuurlijk: wat is informele semantiek?}Consider for example a steel oven scheduling problem. For every block of steel, we should find a time $t$ to put that block in the oven and at time $t+D$, where $D$ is some fixed delay, we take the block out. 
In \clog this is modelled as 
\[
	\Call{b}{Block(b)}{\Csel{t}{\ltrue}{\Cand{In(b,t)}{Out(b,t+D),}    }   }
\]
but to model this in \FO we would get one similar constraint together with several constraints guaranteeing uniqueness:
\begin{align*}
	&\forall b [Block(b)]: \exists t:In(b,t) \land Out(b,t+D)\\
	&\forall b, t, t' [Block(b)]: In(b,t)\land In(b,t')\limplies t= t'\\
	&\forall b, t, t' [Block(b)]: Out(b,t)\land Out(b,t')\limplies t= t'\\
	&\forall x: (\exists t: In(x,t)\lor Out(x,t)) \limplies Block(x)
\end{align*}
Here, the approach in \clog is much more natural, as in this example it is clear how to construct a model, 
whereas to model it in \FO, we should analyse all properties of models.
On the other hand, if we extend this example with a constraint that no two blocks can enter the oven at the same time, this is easily expressible in \FO:
\[
 \lnot \exists t, b, b'[b\neq b']: In(b,t)\land In(b',t),
\]
while this is not naturally expressible in \clog. 
This shows the power of \foclog, the integration of \fo and \clog. 
For example, the entire above scheduling problem would be modelled in \foclog as follows (where we use ``$\{$'' and ``$\}$'' to separate the \clog theory from the \fo sentences).
\begin{ltheo}
 \begin{ldef}
	\Call{b}{Block(b)}{\Csel{t}{\ltrue}{\\\qquad\Cand{In(b,t)}{Out(b,t+D)}    }   }
 \end{ldef}\\
 \lnot \exists t,b, b'[b\neq b']: In(b,t)\land In(b',t)
\end{ltheo}
This is much more readable and much more concise than any pure \clog or \fo expression that expresses the same knowledge. 
As can be seen, the integration of the orthogonal languages \fo and \clog, \foclog provides a great modelling flexibility.

\section{\foclog and ASP}\label{sec:asp}
\marc{deze sectie moet beginnnen met ``D-ASP is not a causal logic. It does not respect the principle of independent causation''}.
\marc{vergelijk EERST met disjunctive-ASP (geen exists) DAARNA met E-disj. Zeg iets over grounding van E-disj naar disj}
\marc{te veel focus op ASP}

The methodology from the previous section is very similar to the ``generate, define, and test'' (GDT) methodology used in Answer Set Programming (ASP). 
In that methodology, ``generate'' and ``define'' are constructive modules of ASP programs that describe which atoms can be  true, while the 
``test'' module corresponds to first-order sentences that constrain solutions. 
In \cite{DeneckerLTV12}, it has been argued that GDT programs correspond to \foid theories. 
Furthermore, in \mycite{clog}, we showed that \foid is syntactically and semantically a sublanguage of \foclog. 
Here, we argue that a more general class of ASP programs can be seen as \foclog theories.

 E-disjunctive programs \cite{tplp/YouZZ13} are finite sets of rules of the form:
\begin{equation}\label{eq:asprule}
	\forall \xxx :\exists \yyy: \alpha_1; \dots;\alpha_m \asprule \beta_1,\dots,\beta_k, \aspnot \gamma_1,\dots,\aspnot \gamma_n.
\end{equation}
where the $\alpha_i, \beta_i$ and $\gamma_i$ are atoms and variables in \yyy only occur in the $\alpha_i$. 
Given a structure $\M$, we define $\M^-$ as the literal set 
\[\{\lnot \alpha\mid \alpha \text{ is a domain  atom on }dom(\M) 
\text{ and }\M\not\models\alpha\}.\]
A structure \M is a  stable model of E-disjunctive program $\PP$ (denoted $\M\models\PP$) if \M is a minimal set $X$ satisfying the condition: 
for any rule $r\in \PP$ and any variable assignment $\eta$, if the literal set 
$X\cup\M^-$ logically entails $body(r)\eta$, then for some assignment $\theta$, 
and for some $\alpha$ in the head of $r$, $(\alpha\eta|_{\xxx})\theta\in X$.
A rule of the form (\ref{eq:asprule}) is called a \emph{constraint} if $m= 0$.

\begin{definition}\label{def:asptrans}
	Let \PP be an E-disjunctive program. The \emph{corresponding} \foclog-theory is the theory $\theory_\PP$ with as \clog expression the \Cand-conjunction of all expressions 
\begin{align*}
&\Call{\xxx}{\beta_1\land\dots\land\lnot\gamma_n}{ \Csel{\yyy}{\ltrue}{\Cor{\alpha_1}{\Cor{\dots}{\alpha_m}}}}
\end{align*}
such that there is a rule of the form (\ref{eq:asprule})
with $m>0$ in \PP.
$\theory_\PP$ has as \fo part:
\begin{itemize}
 \item all sentences 
		$\forall \xxx: \lnot (\beta_1\land\dots\land\beta_k\land\lnot \gamma_1\land\ldots\land\lnot\gamma_n)$
	such that there is a rule of the form (\ref{eq:asprule}) with $m=0$ (i.e., a constraint)
 in \PP and 
  \item 
the sentences $\forall \xxx: \lnot P(\xxx)$
for symbols $P$ that do not occur in the head of any rule in \PP.
\end{itemize}
\end{definition}

The last type of constraint is a technical detail: in ASP, all symbols are \tcaused,  while in \clog, this is only the case for predicates occurring in ``the head of rules''.

The above syntactical correspondence does not always correspond to a semantical correspondence. 
Intuitively, an E-disjunctive rule $r$ (roughly) means the following:
if the body of $r$ holds for an instantiation of $\xxx$, then we select one instantiation of the $\yyy$ and one disjunct; that disjunct is caused to be true for that instantiation. 
But, globally the selection should happen in such a way that the final model is minimal. 
For example the program
$
	\{p.\quad p; q. \}
$
only has one stable model, namely $\{p\}$. 
The intuition behind it is that the first rule causes $p$ to be true, and hence compromises the choice in the second rule. 
As $p$ already holds, the global minimality condition ensures that the second rule is obliged to choose $p$ as well, if possible. When we slightly modify the above program, by adding a constraint:
$
	\{p. \quad p; q. \quad \asprule \aspnot q.\}
$
suddenly, $q$ can (and should) be chosen by the second rule, as $\{p\}$ no longer is a model of this theory.
The above illustrates that there is a great interdependency between different rules and between rules and constraints: adding an extra rule or constraint changes the meaning of other rules. 
Below, we identify a fragment of E-disjunctive ASP in which this dependency is not too strong, and we show that for this fragment, the stable model semantics equals the \foclog semantics. In order to do so, we introduce the following concepts:

\begin{definition}
	Let $\delta$ be a domain atom and $r$ a rule in the form of (\ref{eq:asprule}). Suppose $\eta$ is a variable assignment of the variables $\xxx$ and $\yyy$.
	We say that $\delta$ \emph{occurs in} $r$ at $i$ for $\eta$ if 
	$\alpha_i\eta = \delta$.
	We say that $\delta$ \emph{occurs in} $r$ if there exist and $i$ and an $\eta$ such that $r$ occurs at $i$ for $\eta$.
\end{definition}

\begin{definition}
	We call a rule \emph{disjunctive} if $\yyy$ is not the empty tuple or if $m>1$.
\end{definition}

\begin{definition}
	An E-disjunctive program  $\PP$ is called \emph{non-overlapping} if for every domain atom $\delta$ one of the following holds
\begin{itemize}
	\item $\delta$ occurs only in non-disjunctive rules, or
	\item there are at most one rule $r$, one  $i$, and one $\eta$ such that $\delta$ occurs in $r$ at $i$ for $\eta$.
\end{itemize}
\end{definition}

The above condition  states that domain atoms occurring in heads of disjunctive rules, cannot occur multiple times in rule heads. 
Intuitively, this guarantees that different choices do not interfere. 

\begin{thm}\label{thm:asp}
	Let $\PP$ be a non-overlapping E-disjunctive program without recursion over negation and $\theory_\PP$ the corresponding \foclog theory.
	For every structure $\I$, $\I\models \PP$ if and only if $\I\models\theory_\PP$.
\end{thm}

In Theorem \ref{thm:asp}, there is one extra condition on  non-overlapping ASP programs to be equivalent to the corresponding \foclog theory,
namely that it does not contain recursion over negation, i.e., there are no rules of the form
\begin{align*}
	p\asprule\aspnot p'.\qquad p'\asprule\aspnot p.
\end{align*}
It has already been argued in \cite{DeneckerLTV12} that in practical applications recursion over negation is mostly for two purposes: 1) expressing constraints and 2) to ``open'' the predicate $p$, i.e., to encode that it can have arbitrary truth value. In this case, the predicate $p'$ would not be used in the rest of the theory. 
This can as well be done with a rule 
$
	p;p'.
$
This last rule is equivalent to the above two  in non-overlapping programs (or, if $p$ and $p'$ do not occur in other rule heads). 
In \foclog, we could either add the disjunctive rule, or simply omit this rule, since \topen predicates are open anyway. 

As already stated above, in case an ASP program is not non-overlapping, semantics might differ. However, we do have 
\begin{thm}
	Let $\PP$ be any E-disjunctive program without recursion over negation and $\theory_\PP$ be the corresponding \foclog theory.
	For every structure $\I$, if $\I\models \PP$ then also $\I\models\theory_\PP$.
\end{thm}
The reverse does not hold, since \clog does not impose a global minimality condition. The difference in semantics is illustrated in the American Lottery example, which we resume below.

In the above, we argued that for 
many practical applications of E-disjunctive programs, semantics of \foclog corresponds to the stable model semantics. 
This raises the question of relevance of \foclog. \marc{deze sectie moet hier niet staan. Je moet je dat niet afvragen, als van in het begin duidelijk is dat D-ASP geen causale logica is, stelt deze vraag zich niet}
From a knowledge representation perspective, \foclog adds several useful constructs with respect to E-disjunctive logic programs. 
Among these are nested rules (in fact, arbitrary nesting of expressions), dynamic choice, object creation, and a more modular semantics.

\emph{Nested causal rules} occur in many places, for example,  one could state that the electrician causes a causal link between a button and a light, e.g.,
\[\Cif{electrician}{(\Cif{button}{light})}.\] \marc{niet overtuigend.. Beter $p$ veroorzaakt: als $r$ dan $q$ als niet $r$ dan $s$}
We found similar nested rules in \cite{DBLP:journals/corr/KowalskiS13}.
Of course, for simple examples this can also be expressed compactly in ASP, e.g. by 
\[light\asprule electrician, button.\]
but when causes and effects are more complex, translating them requires the introduction of auxiliary predicates, diminishing the readability of the resulting program.

\emph{Dynamic choices} occur in many practical applications. 
Consider the following situation: a robot enters a room, opens some of the doors in this room, and then leaves by one of the doors that are open. The robot's leaving corresponds to a non-deterministic choice between a {\em dynamic} set of alternatives, which is determined by the robot's own actions, and therefore cannot be hard-coded into the head of a rule. In \clog, we would model this last choice as
\[\Csel{x}{open(x)}{leave(x)}. \]\marc{niet goed want open is hier exo ipv endo zoals de tekst beweert}
To model this in an E-disjunctive logic program, we need an extra auxiliary predicate, thus reducing readability:
\begin{align*}
	&\exists X: chosen(X).\\
	&\forall X: leave(X) \asprule  chosen(X).\\
	&\forall X\asprule  chosen(X); \aspnot open(X).
\end{align*}

\emph{Modularity} of the semantics has already been discussed above: \marc{VEEL EERDER}
The non-overlapping condition on ASP programs guarantees similar modularity. 
However, when the non-overlapping condition is violated, semantics of ASP programs are often less clear. 
Let us reconsider Example \ref{ex:lottery}. 
The E-disjunctive program
  \begin{align*}
  \exists X&: permres(X)\asprule lottery.\\
   \forall X&: permres(X) \asprule passtest(X).
 \end{align*}
is similar to  \begin{align*} \begin{ldef} &\Cif{lottery}{\left(\Csel{x}{\Tr}{permres(x)\right)}}\\ &\Call{x}{passtest(x)}{permres(x)}\end{ldef}
 \end{align*}
Semantically, the first imposes a minimality condition: the lottery is always won by a person succeeding the test, if there exists one.  
On the other hand, in \foidplus the two rules are independent, and models might not be minimal. 
In this example, it is the latter that is intended. 
This illustrates modularity of \clog. The rule $\Cif{lottery}{\left(\Csel{x}{\Tr}{permres(x)\right)}}$ means that one person is selected randomly to obtain residence. 
Adding other rules does not change the meaning of this rule; causal effects do not interfere.


\emph{Object-creation} in \clog is discussed in the next section.

\section{Object-creation in \clog} \label{sec:new}
Object creation is available in \clog  through the \Cnew-operator. 
Like every language construct in \clog, the informal interpretation of an expression 
\[
 \Cnew{x}{P(x)}\lrule\varphi
\]
is defined in terms of causal processes. 
The above expression states that $\varphi$ causes the creation of a new element and that for that new element, $P$ is caused. 
Object-creation is also subject to the principles of sufficient causation, universal causation and no self-causation.
In order to apply these principles, the domain of a structure is partitioned into two parts: the \emph{initial} elements are those whose existence is
not governed by the causal theory, they are \topen and the \emph{created} elements are those created by expressions in the causal theory, i.e., they are \tcaused. 
For created elements, their default value is \emph{not existing} and their deviant value is existing.
Thus, at the start of a causal process, only the initial elements exist, as soon as the preconditions of a \Cnew-expressions are satisfied, an element is added to the domain. 
The principle of no self-causation takes these default and deviant values into account: an object cannot be created based on its own existence.
Consider for example the following causal theory:
\begin{align*}
&\Csel{x}{\ltrue}{P(x)}\\
 &\Cif{\exists x: P(x)}{(\Cnew{y}{Q(y))}}\\
&\Csel{x}{\ltrue}{R(x)}
 \end{align*}
 The first and last expressions select one object randomly and cause $P$ (respectively $R$) to hold for that object. 
 The second expression creates a new element conditionally, only if there is at least one element satisfying $P$. 
In this example, the element selected for the first expression cannot be the one created in the second. 
$\Csel$-operators can only select existing elements and the object created in the second expression can only be created after the selection in the first rule, after there is some object satisfying $P$.
For the last expression, any element can be selected. 
Hence, this causal theory has no models with only one domain element. 
A structure \I with domain $\{A,B\}$ and with $P^\I=\{A\}$ and $Q^\I=R^\I=\{B\}$ is a model of the above causal theory. 
In this case, $B$ is the unique created element, and $A$ is initial, i.e., $A$ is assumed to exist before the described causal process  takes place.
This illustrates that the $\Cnew$-operator is more than simply a \Csel together with unique name axioms:
its semantics is really integrated in the underlying causal process. 
The behaviour of \Cnew-expressions can be simulated using \Csel-expressions if we make the two parts of the domain (initial and created elements) explicit and conditionalise all quantifications. A detailed discussion of this transformation is out of the scope of this paper.

Object creation occurs in many fields, of which we discuss some below.

\subsection{Object-Creation in Database Systems}
Object-creation has been studied intensively in the field of deductive databases. 
In \cite{jcss/AbiteboulV91}, various extensions of Datalog, are considered, resulting in non-deterministic semantics for queries and updates. 
One of the studied extensions is object creation (throught existential quantifications in rule heads). 
These and similar related extension have been implemented in several systems, including LogicBlox \cite{datalog/GreenAK12}.  
An example from the latter paper is the rule: 
\todo{p. 6.: Instead of simply using math mode, use e.g. $\mathit{presidentOf}$.}
\[ President(p), presidentOf[c] = p \lrule Country(c).\]
which means that for every country $c$, a new (anonymous) ``derived
entity'' of type $President$ is created. 
Of course, the president of a country is not a new person, but the president is new with respect to the database, which does not contain any persons yet. 
Such rules with (implicit) existentially
quantified head variables correspond to $\Cnew$-expressions. Here, it would translate to
\[\Call{c}{Country(c)}{\Cnew{p}{\Cand{Pres(p)}{presOf(c,p)}}}.\]
This shows that in some rule-based paradigms, an existentially quantified head-variable corresponds to object-creation (\Cnew), 
while in other rule-based paradigms, such as ASP, we saw that an existentially quantified head variable corresponds to a selection. 
The relation between these paradigms has, to the best of our knowledge, not yet been studied thoroughly. 
We believe that \foclog, which makes an explicit distinction between selection and object-creation, is an interesting tool to study this relationship. 
This is future work.

Many other Datalog extensions with forms of object creation exist. 
For example \cite{VandenBusscheP95} discusses a
version with creation of sets and compares its expressivity 
with simple object creation. 

Object-creation also occurs in other database languages, such as for example the query language $\mathrm{while_{new}}$ in \cite{aw/AbiteboulHV95}.
An expression 
\[\texttt{\tt while R do  (P = new Q)}\]
 in that language   corresponds to a \CEE.
\[\Call{t}{R(t)}{\Call{\xxx}{\ltrue}{\Cnew{y}{ P(\xxx,y,t+1)} \lrule Q(\xxx,t)}.}\]
In fact in \mycite{inferenceClog}, it has been shown that \clog can ``simulate'' the entire language  $\mathrm{while_{new}}$.

\subsection{Object-Creation in Mathematics}
\marc{ik zou dit voor de databases zetten...}
Object-creation also occurs in mathematics. 
The set of all natural numbers can be thought of as the set obtained by a process that first creates one element (zero) and for every element in this set, adds another element (its successor). 
In \clog, the above natural language sentences can be modelled as follows
\begin{align*}
&\Cnew{x}{ (\Cand{\pred{Nat}(x)}{\pred{Zero}(x)})}\\
&\Call{ x}{\pred{Nat}(x)}{\Cnew{y}{ (\Cand{\pred{Nat}(y)}{\pred{Succ}(x,y)}).}}
\end{align*}
Models of the above theory are exactly those structures interpreting $\pred{Nat}, \pred{Zero}, \pred{Succ}$ as the natural numbers, zero and the successor function (modulo isomorphism).

\subsection{Object-Creation in Business Rules Systems}
Business Rules \cite{Group2000Defining} engines are widely used in the industry. 
One big drawback of these systems is their inability to perform multiple forms of reasoning. 
For example, banks might use a  Business Rules engine to decide whether someone is  eligible for a loan. 
This approach can be very efficient, but as soon as one is not only interested in the above question, but also in explanations, 
or suggestions about what to change in order to become eligible, the application should be redesigned. 
Previous attempts to translate Business Rules applications into a logic with a Tarskian model semantics have been made in \cite{iclp/VanHertum13}. The conclusion of this study was that for such a transformation, we need object creation . 
We believe that \clog provides a suitable form of object-creation for this purpose. 
As an illustration, the JBoss manual \cite{browne2009jboss} contains the following rule: 
 \begin{lstlisting}
 when Order( customer == null )
 then insertLogical(new ValidationResult(
   “validation.customer.missing”));\end{lstlisting}
This rule means that if an order is created without customer, a new \textit{ValidationResult} is created with the message that the customer is missing. This can be translated to \clog as follows:
\begin{align*}
	&\Call{y}{\pred{Order}(y)\land\pred{NoCustumer}(y)}{\\
	 & \quad\Cnew{x}{\Cand{\pred{ValidationR}(x)}{\pred{Message}(x,\text{``\dots''})}}}.
\end{align*}
A more thorough study of the relationship between the operational semantics of Business Rules systems and the semantics of \clog is a topic for future work.

\todo{The discussion on object creation might be extended by a comparison
with the transformation rules used in Abstract State Machines which
also have an explicit new-element construct.}

\todo{I think recently there has been renewed general interest in general in
value creation in Datalog, not only in LogicBlox, but also in
Datalog+-, Datalog-exists, existential rules, and similar formalisms,
often used in the realm of ontologies, where "value-generating"
existentials are commonplace. It would be a good idea to extend the
comparison in that direction as well}

\todo{One lack that I noted was that there are many recent approaches to value generation in Datalog and other rule-like languages, for instance Datalog+/-, or existential rules, or rules in ontology settings, that would be relevant to this work. }

\bart{evt nog extra ref toevoegen naar andere datalog dingen}

\section{Conclusion}\label{sec:concl}
In this paper we compared \foclog to other modelling paradigms. 
We discussed the semantical relationship between \clog and \fo.
We identified a fragment of E-disjunctive logic programs for which the  stable model semantics corresponds to the semantics of \foclog, and argued how \foclog enriches such programs with several useful modelling constructs.
Furthermore, we argued that the object-creation in \foclog corresponds to the object creation in many related language. 
Besides technical relationship between these languages, we believe that this discussion also provides insights in the semantics of \foclog.

 \bibliography{krrlib}

\end{document}